\RequirePackage{ifpdf}
\ifpdf 
\documentclass[pdftex]{sigma}
\else
\documentclass{sigma}
\fi

\usepackage{fancybox}
\usepackage[all]{xy}

\numberwithin{equation}{section}

\def\p{\partial}

\def\real{\mathbb{R}}
\def\complex{\mathbb{C}}
\def\act{\triangleright}
\def\and{\quad{\rm and}\quad}

\newcommand{\Bra}[1]{\left\langle\, #1\,\right|}
\newcommand{\Ket}[1]{\left|\, #1\,\right\rangle}

\def\CA{{\cal A}}   \def\CD{{\cal D}}
 \def\CF{{\cal F}}  \def\CH{{\cal H}}
   
 \def\CN{{\cal N}} \def\CO{{\cal O}} \def\CP{{\cal P}}

\begin{document}

\allowdisplaybreaks

\renewcommand{\thefootnote}{$\star$}

\renewcommand{\PaperNumber}{068}

\FirstPageHeading

\ShortArticleName{Twist Quantization of String and Hopf Algebraic Symmetry}

\ArticleName{Twist Quantization of String\\ and Hopf Algebraic Symmetry\footnote{This paper is a
contribution to the Special Issue ``Noncommutative Spaces and Fields''. The
full collection is available at
\href{http://www.emis.de/journals/SIGMA/noncommutative.html}{http://www.emis.de/journals/SIGMA/noncommutative.html}}}

\Author{Tsuguhiko ASAKAWA and Satoshi WATAMURA}
\AuthorNameForHeading{T.~Asakawa and S.~Watamura}
\Address{Department of Physics,
Graduate School of Science,
Tohoku University,\\
Aoba-ku, Sendai 980-8578, Japan}
\Email{\href{mailto:asakawa@tuhep.phys.tohoku.ac.jp}{asakawa@tuhep.phys.tohoku.ac.jp}, \href{mailto:watamura@tuhep.phys.tohoku.ac.jp}{watamura@tuhep.phys.tohoku.ac.jp}}

\ArticleDates{Received April 07, 2010, in f\/inal form August 12, 2010;  Published online August 20, 2010}

\Abstract{We describe the twist quantization of string worldsheet theory,
which unif\/ies the description of quantization and the target space symmetry,
based on the twisting of Hopf and module algebras.
We formulate a method of decomposing a twist into successive twists
to analyze the twisted Hopf and module algebra structure,
and apply it to several examples, including f\/inite twisted dif\/feomorphism
and extra treatment for zero modes.}

\Keywords{string theory; qunatization; Hopf algebra; Drinfeld twist}

\Classification{83E30; 81T75; 53D55}

\renewcommand{\thefootnote}{\arabic{footnote}}
\setcounter{footnote}{0}

\section{Introduction}
String theory is a promising candidate as a unif\/ied theory of
elementary particles and quantum gravity.
There are many evidences that string theory contains quantum gravity,
the existence of graviton in the spectrum for example,
and the fact that the conformal symmetry
in the non-linear sigma-model leads to the Einstein equation.
However, it is not yet clear how the concepts of general relativity,
such as general covariance or equivalence principle,
are realized in string theory.
The origin of these dif\/f\/iculties is that the theory is formulated only
as a f\/irst quantized theory in a f\/ixed background,
and it is dif\/f\/icult to consider a change to other backgrounds.

In \cite{AMW} and \cite{AMW2},
we proposed a new formulation of string worldsheet theory
based on the Hopf algebra structure which is minimally background dependent.
In~\cite{AMW},
we gave a unif\/ied formulation of the quantization of the string and the
spacetime symmetry, by reformulating
the path-integral quantization of the string as a
Drinfeld twist~\cite{Drinfeld:1990qh} of the Hopf algebra.
By the twist, the space-time dif\/feomorphism was deformed
into a twisted Hopf algebra, while the Poincar\'e symmetry was
kept unchanged.
The method was applied to the constant $B$-f\/ield background in~\cite{AMW2}
and the structure of the twisted Hopf algebra in that case was studied.
As a quantization, it implies a new normal ordering,
 and as a symmetry,
we have shown that the twisted Poincar\'e symmetry~\cite{Chaichian:2004za,Koch:2004ud,Aschieri:2005yw}
in the noncommutative
f\/ield theory on the Moyal plane are derived from our twisted Hopf algebra
structure.

One of purposes of this paper is to give a overview on the twist quantization
developed in \cite{AMW,AMW2} so far. Part of the paper's results was previously presented at the workshop
on Quantum Gravity and Quantum Geometry (Corfu, 2009) and published in its proceedings~\cite{SW}.
We repeat this material here for completeness' sake and for readers'
convenience.
We then formulate the method of decomposing a twist in a more general
setting by extending \cite{AMW2},
and discuss that it has wide variety of applications.
New results of this paper include an improvement for interpretation of the twisted dif\/feomorphism
from the viewpoint of vertex operators,
and the treatment of the zero modes for momentum conserving vacuum.

The paper is organized as follows:
In Sections~\ref{sec:A} and \ref{sec:twist quantization}, we give a brief review of the twist quantization
of string theory \cite{AMW}.
In Section~\ref{sec:relating}, we describe a method of splitting the twist in great generality.
Then, in Section~\ref{section5}, we give three examples applying the method,
including the case for a constant $B$-f\/ield background \cite{AMW2} and
two new examples.
Section~\ref{section6} is devoted to discussion and conclusion.

\section{Classical Hopf algebra in string theory}\label{sec:A}

We start with describing the classical Hopf algebra $\CH$ of
functional derivatives and its module algebra $\CA$ of string functionals \cite{AMW}.
They depend on a target space manifold $M$ (here we f\/ix $M=\real^d$)
but are independent of the background data such as
the metric and $B$ f\/ield.

The basic variable of the string worldsheet theory
is a set of functions $X^\mu(z)$ $(\mu=0,\dots, d-1)$,
or equivalently, a map $X:\Sigma \rightarrow M$ from
the worldsheet $\Sigma$ into a target space $M$.
We denote the space of such maps as ${\cal X}$.
Let $\CA=C({\cal X})$ be the vector space of functionals
$\CA \ni I: X \mapsto I[X] \in \complex$
of the abstract form
\begin{gather}
I[X]= \int  d^2 z\, \rho(z) F[X(z)] ,
\label{def of functional}
\end{gather}
where $F[X(z)] \in C^\infty(\Sigma)$
is given from a tensor f\/ield in the target space by pull-back,
and $\rho(z)$ is some weight function (distribution).
In particular, we call
\begin{gather}
F[X](z_i)= \int  d^2 z\, \delta^{(2)}(z-z_i) F[X(z)]  ,
\label{local functional}
\end{gather}
a local functional at $z_i$, with an additional label $z_i \in \Sigma$.
These functionals~(\ref{def of functional}) and~(\ref{local functional})
correspond to an integrated vertex operator
and to a local vertex operator after quantization, respectively.
For example, for a one-form $\Omega=\Omega_\mu(x)dx^\mu$ in the target space
we can def\/ine a functional
\begin{gather}
I_\Omega [X] = \int  d^2 z\, \rho^a (z) \p_a X^\mu \Omega_\mu [X(z)]  .
\label{1 from functional}
\end{gather}
$\CA$ forms a commutative algebra over $\complex$,
by def\/ining a multiplication of two functionals
$m : \CA\otimes \CA \rightarrow \CA$ as
$(IJ)[X]=I[X]J[X]$.

Let ${\mathfrak X}$ be the space of all functional vector f\/ields of the form
\begin{gather}
\label{gene vector field}
\xi = \int   d^2w \, \xi^\mu (w) \frac{\delta}{\delta X^{\mu}(w)},
\end{gather}
and denote its action on $I[X] \in \CA$ as $\xi \act I[X]$.
Here $\xi^\mu (w)$ is a weight function
(distribution) on the worldsheet including the following two classes.
\begin{itemize}\itemsep=0pt
\item[i)] $\xi^\mu(w)$ is a  pull-back of a target space function
$\xi^\mu (w)= \xi^\mu [X(w)]$.
It is related to the variation of the functional under dif\/feomorphism
$X^\mu(z) \rightarrow X^\mu(z) +\xi^\mu[X(z)]$ as
$\delta_\xi F[X] = -\xi\act F[X]$.
Note that it is arranged so that
$(\xi \act I_\Omega) [X] = I_{{\cal L}_\xi \Omega}[X]$
for a functional (\ref{1 from functional}),
where ${\cal L}_\xi$ is the Lie derivative acting on the target space one-form $\Omega$.
\item [ii)] $\xi^\mu (w)$ is a function of $w$ but independent of $X(w)$ and its derivatives.
It corresponds to a~change of the embedding,
$X^\mu(z) \rightarrow X^\mu(z) +\xi^\mu(z)$, and is
used to derive the equation of motion.
\end{itemize}
By successive transformations, $\xi \triangleright (\eta \triangleright F)$,
the vector f\/ields $\xi$ and $\eta$
form a Lie algebra with the Lie bracket
\begin{gather*}
\label{Lie algebra}
[\xi, \eta]= \int   d^2w   \left(
\xi^\mu \frac{\delta \eta^\nu}{\delta X^{\mu}}
-\eta^\mu \frac{\delta \xi^\nu}{\delta X^{\mu}}
\right) (w) \frac{\delta}{\delta X^{\nu}(w)} .
\end{gather*}
Once we have a Lie algebra ${\mathfrak X}$, we can def\/ine its universal
enveloping algebra $\CH=U({\mathfrak X})$ over $\complex$,
which has a natural cocommutative Hopf algebra structure
$(U({\mathfrak X}); \mu,\iota,\Delta,\epsilon,S)$\footnote{A similar Hopf algebra based on functional derivatives is also considered in \cite{Aschieri}.}.
The def\/ining maps given on elements $k\in\complex$ and
$\xi, \eta \in {\mathfrak X}$ are
\begin{gather}
  \mu (\xi \otimes \eta )=\xi \cdot \eta,\qquad
\Delta(\xi )=\xi  \otimes 1+1 \otimes \xi,\qquad
\epsilon(\xi)=0,\qquad S(\xi )=-\xi,\nonumber\\
 \iota(k)=k\cdot 1,\qquad
\Delta(1)=1\otimes 1,\qquad
\epsilon(1)=1,\qquad S(1)=1.
\label{defing maps}
\end{gather}
As usual, these maps are uniquely extended to any element of $U({\mathfrak X})$ by the
algebra (anti-) homomorphism.
There are many subalgebras in ${\mathfrak X}$.
In particular, the Poincar\'e--Lie algebra $\CP$ generated by
\begin{gather}
P^{\mu}=-i \int d^2z \,\eta^{\mu\lambda}
\frac{\delta}{\delta X^{\lambda}(z)},
\qquad L^{\mu\nu}=-i \int d^2z  X^{[\mu}(z) \eta^{\nu]\lambda}
\frac{\delta}{\delta X^{\lambda}(z)},
\label{generators}
\end{gather}
where $P^\mu$ are the generators of the translation and $L^{\mu\nu}$ are
the Lorentz generators,
forms a Hopf subalgebra of $\CH=U({\mathfrak X})$, denoted as $U(\CP)$.
We denote the Abelian Lie subalgebra consisting of~$\xi$~(\ref{gene vector field}) in class ii)
as ${\mathfrak C}$.
Then $U({\mathfrak C})$ is also a Hopf subalgebra of $\CH$.

The algebra $\CA$ of functionals is now considered as a
$\CH$-module algebra.
The action on the product of two elements $F,G \in \CA$ is a natural generalization of the action of $\xi$, def\/ined by
\begin{gather*}
h \triangleright m (F\otimes G) = m \Delta(h) \triangleright (F\otimes G) ,
\label{covariance}
\end{gather*}
satisfying the Leibniz rule ($\Delta (\xi)$ in (\ref{defing maps})).

\section{Twist quantization}
\label{sec:twist quantization}

In~\cite{AMW} we gave a simple quantization procedure in terms of a
Hopf algebra twist,
in which the vacuum expectation value (VEV) coincides with the conventional
path integral average.
A~general theory of Hopf algebra twist is presented in
\cite{Drinfeld:1990qh,2000fqgt.book.....M}.

Suppose that there is a twist element
(counital 2-cocycle), $\CF \in \CH\otimes \CH$, which is
invertible, counital $({\rm id} \otimes \epsilon)\CF=1$
and satisf\/ies the 2-cocycle condition
\begin{gather}
 (\CF \otimes {\rm id})(\Delta \otimes {\rm id})\CF=({\rm id}\otimes\CF)
({\rm id} \otimes \Delta)\CF.
\label{2-cocycle condition}
\end{gather}
Given a twist element $\CF$, the twisted Hopf algebra $\CH_{\CF}$ can be
def\/ined by modifying the coproduct and antipode as
\begin{gather*}
\Delta_{\CF}(h)=\CF \Delta (h) \CF^{-1}, \qquad
S_{\CF}(h)=US(h)U^{-1}
\end{gather*}
for all $h \in \CH$, where $U=\mu ({\rm id}\otimes S)\CF$.
We regard this procedure of twisting as
a quantization with respect to the twist element~$\CF$.

Correspondingly, the consistency of the action, i.e.\ covariance with
respect to the Hopf algebra action,
requires that a $\CH$-module algebra~$\CA$ is twisted
to the $\CH_{\CF}$-module algebra~$\CA_{\CF}$.
As a vector space, $\CA_{\CF}$ is identical to~$\CA$,
but is accompanied by the twisted product
\begin{gather*}
F*_{\CF} G =m_{\CF} (F\otimes G) :=m\circ \CF^{-1}\act (F\otimes G) ,
\label{twistedProduct}
\end{gather*}
which is associative owing to
the cocycle condition (\ref{2-cocycle condition}).
The covariance means that for $h\in \CH_{\CF}$ and $F,G \in \CA_{\CF}$,
\begin{gather*}
 h\act m_{\CF} (F\otimes G)
 =  m\circ \Delta (h) \CF^{-1}\act (F\otimes G)
 =  m_{\CF} \Delta_{\CF} (h) \act (F\otimes G) .
\label{twisted covariance}
\end{gather*}
In this way the Hopf algebra and the module algebra are twisted in a consistent manner.
The resulting twisted action is considered as a transformation
in the quantized theories.

We def\/ine the vacuum expectation value (VEV) for the twisted module algebra $\CA_{\CF}$
as the map $\tau: \CA_{\CF} \rightarrow \complex$.
For any element $I[X] \in \CA_{\CF}$ it is def\/ined as the evaluation at $X=0$:
\begin{gather}
\tau\left( I[X]  \right):=I[X]|_{X=0}.
\label{def of true VEV}
\end{gather}
If $I[X]$ is a product of two elements $F,G\in \CA_{\CF}$,
it gives a correlation function
\begin{gather*}
\sigma(z,w)=\tau(F[X(z)]*_{\CF}G[X(w)])
=\tau\circ m\circ \CF^{-1} \act (F\otimes G).
\end{gather*}
Owing to the associativity,
the correlation function of $n$ local functionals is similarly given as
\begin{gather*}
\sigma(z_1,\dots,z_n)=\tau(F_1[X(z_1)]*_{\CF}F_2[X(z_2)]\cdots *_{\CF}F_n[X(z_n)]).
\end{gather*}
The action of $h\in\CH_{\CF}$ inside the VEV $\tau (I[X])$ is
\begin{gather*}
\tau \left(h \act I[X]\right) .
\label{Hopf action}
\end{gather*}
This action appears in the various relations related with the symmetry transformation.

This completes the description of the procedure.
A dif\/ferent choice of a twist element $\CF$ gives a quantization
in a dif\/ferent background.

\subsection{Normal ordering and path integrals}
\label{sec:normal ordering}

Here we assume that a twist element
is in the Abelian Hopf subalgebra $\CF \in U({\mathfrak C})\otimes U({\mathfrak C})$
of the form
\begin{gather}
\CF=\exp\left(-{\int  d^2z  \int d^2w \,
G^{\mu\nu}(z,w)
\frac{\delta}{\delta X^{\mu}(z)}
\otimes\frac{\delta}{\delta X^{\nu}(w)}}\right),
\label{eq:F}
\end{gather}
specif\/ied by a Green function $G^{\mu\nu}(z,w)$ on the worldsheet $\Sigma$.
It is easy to show that this~$\CF$ in~(\ref{eq:F}) satisf\/ies all conditions
for twist elements~\cite{AMW}.
In this case, the twist quantization described
in the previous subsection is identical with
the path integral quantization.

As proved in \cite{AMW}, the twist element $\CF$ in (\ref{eq:F})
can be written as
\begin{gather}
 \CF =\big(\CN ^{-1}\otimes \CN ^{-1}\big)\Delta (\CN ),
\label{coboundaryrelation}
\end{gather}
where the normal ordering element $\CN \in \CH$ is def\/ined by
\begin{gather}
\CN  = \exp\left\{-\frac{1}{2}\int  d^2z  \int  d^2w \,
G^{\mu\nu}(z,w)
\frac{\delta}{\delta X^\mu(z)}\frac{\delta}{\delta X^\nu(w)}\right\}  .
\label{normalorderingoperator}
\end{gather}
This shows that the twist element $\CF$ is a coboundary
and thus it is trivial in the Hopf algebra cohomology.
Consequently, there are isomorphisms summarized as:
\begin{gather}
\begin{array}{@{}ccccc}
\CH & \xrightarrow{\mbox{twist}} & \CH_{\CF} & \xrightarrow{\sim} &
\hat{\CH} \\
\triangledown & & \triangledown & & \triangledown  \\
\CA & \xrightarrow{\mbox{twist}} & \CA_{\CF} & \xrightarrow{\sim}
& \hat{\CA}
\end{array}
\label{diagram}
\end{gather}
In the diagram, the left column is the classical pair $(\CH,\CA)$,
and the middle and the right column are the quantum counterparts.
Here the map $\CH_{\CF} \xrightarrow{\sim} \hat{\CH}$ is given by an
inner automorphism $h \mapsto \CN  h {\CN }^{-1} \equiv \tilde{h}$, and
the map $\CA_{\CF} \xrightarrow{\sim} \hat{\CA}$ is given by
$F \mapsto \CN  \act F \equiv :\!F\!:$.
We call $\hat{\CH}$ ($\hat{\CA}$) the normal
ordered Hopf algebra (module algebra), respectively.

The normal ordered module algebra $\hat{\CA}$ is the one
which appears inside the VEV in the path integral.
It consists of elements in normal ordered form
\begin{gather*}
\CN  \act I[X] \equiv :\!I[X]\!:
\label{normal ordered functional}
\end{gather*}
for any functional $I[X] \in \CA_{\CF}$.
These elements correspond to vertex operators in the path integral,
because the action of $\CN$ (\ref{normalorderingoperator}) corresponds to
subtractions of divergences at coincident points caused by self-contractions
in the path integral.
As an algebra, $\hat{\CA}$ has the same multiplication
$m:\hat{\CA}\otimes\hat{\CA}\rightarrow \hat{\CA}$
as the classical functional $\CA$.
This is seen by the map (\ref{diagram}) of the product in $\CA_{\CF}$ as
\begin{gather*}
 \CN  \act m \circ \CF^{-1} \act (F\otimes G)
= m \circ (\CN  \otimes \CN )\act (F\otimes G) ,
\label{product iso}
\end{gather*}
which is a direct consequence of the coboundary relation (\ref{coboundaryrelation}).
An equivalent but more familiar expression
is the time ordered product of two vertex operators
as $:\! (F *_{\CF} G) \!: = :\!F\!:\, :\!G\!:$.

The normal ordered Hopf algebra $\hat{\CH}$ is def\/ined with the same
algebraic operations as the classical Hopf algebra $\CH$, but
with elements dressed with a normal ordering as
\begin{gather*}
\CN  h {\CN }^{-1} \equiv \tilde{h}
\label{normal ordered h}
\end{gather*}
for any $h \in \CH_{\CF}$.
The isomorphism map (\ref{diagram})
relates the twisted Hopf algebra action $\CH_{\CF}$ on
$\CA_{\CF}$ to the corresponding action of $\hat{\CH}$ on $\hat{\CA}$.
For example,
\begin{gather}
h\act F  \ \xrightarrow{\sim} \
\CN  \act (h \act F) =
\tilde{h}~\act :\!\!F\!:, \label{action on :F:2}\\
h\act (F*_{\CF} G) \ \xrightarrow{\sim}\tilde{h} \ \act (:\!F\!::\!G\!:).
\label{action on :F:}
\end{gather}

The elements in $\hat{\CA}$
contain generically formal divergent series
(this is the reason why we distinguish $\CA$ and $\hat{\CA}$),
and thus equations~(\ref{action on :F:2}) and (\ref{action on :F:}) have only a meaning inside the VEV~\cite{AMW}.
The VEV (\ref{def of true VEV}) for $\CA_{\CF}$ implies the
def\/inition of the VEV
for $\hat{\CA}$ to be a map, $\tau \circ \CN ^{-1}: \hat{\CA}\rightarrow \complex$,
and it turns out that it coincides with the VEV $\langle \cdots \rangle$
in the path integral
\begin{gather}
\langle \CO\rangle :=
\frac{\int \CD X \CO e^{-S}}{\int \CD X e^{-S} } .
\label{VEV}
\end{gather}
For instance, the correlation of two local functionals is
\begin{gather}
\tau \circ \CN ^{-1} \act (:\!F[X]\!:\!(z) :\!G[X]\!:\!(w))
=\langle  :\!F[X]\!:\!(z) :\!G[X]\!:\!(w)  \rangle
\label{VEVpathintegral}
\end{gather}
The equality (\ref{VEVpathintegral}) can be easily
verif\/ied using the standard path integral argument (see for example \cite{Polchinski}).
Clearly, the action functional~$S[X]$ in~(\ref{VEV}) is related to the
Green function in~$\CF$ and~$\CN$.
It is quadratic $S[X]=\frac{1}{2}\int d^2z X^\mu D_{\mu\nu} X^\nu$
with a f\/ixed second order derivative $D_{\mu\nu}$ such that
the Green function in (\ref{normalorderingoperator}) is a solution of
the equation
$D_{\mu\rho}G^{\rho\nu}(z,w) =\delta_\mu^\nu \delta^{(2)}(z-w)$.
This explains why we call a twist element $\CF$ a background
in string worldsheet theory.
Here, the action $\CN ^{-1}$ gives the Wick contraction with respect to the Green function.
Note that in~(\ref{VEV}), each local insertion is understood as being
regulated
by the (conformal) normal ordering~$\CN$.

The two descriptions in terms of a twisted pair $(\CH_{\CF}, \CA_{\CF})$
and
of a normal ordered pair $(\hat{\CH},\hat{\CA})$ (thus path integral) are
equivalent.
However, note that the background dependences in the
two formulations are dif\/ferent.
In the case of the normal ordered pair $(\hat{\CH}, \hat{\CA})$,
both an element $:\!F\!:\in \hat{\CA}$ and the VEV $\tau \circ \CN ^{-1}$
contain $\CN$ which depends on the background.
In the operator formulation this corresponds to the property that
a mode expansion of the string variable $X^\mu(z)$ as well as
the oscillator vacuum are background dependent.
Therefore, the description of the quantization
that makes $\hat{\CA}$ well-def\/ined is
only applicable to that particular background and we need a
dif\/ferent mode expansion for a dif\/ferent background.

\subsection{Twisted Hopf algebraic symmetry}
\label{sec:twisted sym}

One advantage of the twist quantization
is that both, the module algebra $\CA$ (observables)
and the Hopf algebra $\CH$ (symmetry) are simultaneously quantized
by a single twist element $\CF$ (background).
This gives us a good understanding of the symmetry structure
after the quantization,
as discussed in \cite{AMW} in great detail:
There, we argued that the symmetry
of the theory in the conventional sense,
namely the background preserving dif\/feomorphism,
is characterized as a twist invariant Hopf subalgebra of $\CH_{\CF}$,
which consists of elements $h \in \CH$ such that $\Delta_{\CF}(h)=\Delta(h)$
and $S_{\CF}(h)=S(h)$.
Of course, this subalgebra depends on the choice of $\CF$ (background).
The other elements of $\CH_{\CF}$, i.e., generic
dif\/feomorphisms, should be twisted
at the quantum level.

To be more precise, let us f\/ix a twist element to be
the Minkowski background with the metric $\eta_{\mu\nu}$.
We denote it as $\CF_0$.
It is achieved by setting the Green function in (\ref{eq:F}) for that background.
For instance, on the worldsheet $\Sigma=\complex$, it is given by
\begin{gather}
G_0^{\mu\nu}(z,w)  = -{\alpha'}
\eta^{\mu\nu}\ln |z-w| .
\label{Cpropagator}
\end{gather}
In this case, the universal enveloping algebra $U(\CP)$
(\ref{generators}) for the Poincar\'e Lie algebra $\CP$ is the true (quantum) symmetry
of the theory.
It is easily checked that the twist does not alter the
coproduct $\Delta_{\CF_0} (u)=\Delta (u)$
nor the antipode $S_{\CF_0} (u)=S(u)$
for $ \forall \, u \in U(\CP)$ and thus $U({\cal P})$
is identical with the original $U(\CP)$ under the twisting by $\CF_0$.
Therefore, the twist quantization coincides with the ordinary quantization,
in which the Poincar\'e covariance is assumed to hold at the quantum level.

We have not yet fully understood the structure of the twisted dif\/feomorphism
around this background, in particular from the target-space viewpoint.
But it should contain the information on the broken dif\/feomorphism.
There is an argument that
the action of this twisted dif\/feomorphism can be
regarded as a remnant of the classical dif\/feomorphism,
where the change of the background under dif\/feomorphisms is incorporated
into the twisted dif\/feomorphism in such a way that
the twist element (quantization scheme) is kept invariant.
A similar statement in the case of twisted dif\/feomorphism on the Moyal plane
is found in~\cite{AlvarezGaume:2006bn}.
We will come back to this point below.

\section{Decomposition of twists}\label{sec:relating}

In this section, we extend the method discussed in \cite{AMW2}
and describe a general theory
of decomposing a twist into successive twists,
which is a useful tool to relate two dif\/ferent backgrounds.

We will consider two dif\/ferent twist elements $\CF_0$ and $\CF_1$ in $\CH$,
that are related by
\begin{gather}
\CF_1=\CF_d \CF_0~
\label{F_1=F_BF_0}
\end{gather}
for some $\CF_d$.
We assume that the f\/irst twist $\CF_0$ is given by the propagator
$G_0$ as (\ref{eq:F}), which def\/ines a quantization scheme (background)
as described in the previous section.
In general, for a given twisted Hopf algebra $\CH_{\CF_0}$ with a
coproduct $\Delta_{\CF_0}$,
a further twisting by $\CF_d$ is possible if~$\CF_d$ is a twist element in $\CH_{\CF_0}$ (not in $\CH$)
\begin{gather}
(\CF_d \otimes {\rm id})(\Delta_{\CF_0} \otimes {\rm id})\CF_d =({\rm id}\otimes\CF_d)
({\rm id} \otimes \Delta_{\CF_0})\CF_d~.
\label{2nd cocycle condition}
\end{gather}
Then, $\CF_1 =\CF_d \CF_0$ is also a twist element in $\CH$.
This is because the l.h.s.\ of the 2-cocycle condition
(\ref{2-cocycle condition}) for $\CF_1$
can be written as
\begin{gather*}
(\CF_1 \otimes {\rm id})(\Delta \otimes {\rm id})\CF_1
=\left[(\CF_d \otimes {\rm id})(\Delta_{\CF_0} \otimes {\rm id})\CF_d\right]
 \left[(\CF_0 \otimes {\rm id})(\Delta \otimes {\rm id})\CF_0 \right],
\end{gather*}
and similarly for the r.h.s.,
where (\ref{2-cocycle condition}) for $\CF_0$ is used.
Then by using~(\ref{2nd cocycle condition}),
it is shown that $\CF_1$ satisf\/ies~(\ref{2-cocycle condition}).
Conversely, if two twist elements $\CF_0$ and $\CF_1$ in $\CH$ have a relation
$\CF_1=\CF_d \CF_0$,
then $\CF_d$ is a twist element in $\CH_{\CF_0}$ satisfying (\ref{2nd cocycle condition}).
In fact, the decomposition (\ref{F_1=F_BF_0}) def\/ines two successive twists
of the Hopf algebra $\CH$ and the module algebra $\CA$:
\begin{gather}
(\CH,\CA) \  \xrightarrow{\text{twist by $\CF_0$}} \  (\CH_{\CF_0},\CA_{\CF_0})
\  \xrightarrow{\text{twist by $\CF_d$}} \  (\CH_{\CF_1},\CA_{\CF_1}).
\label{2step twist}
\end{gather}
This means that the identities
$(\Delta_{\CF_0})_{\CF_d}=\Delta_{\CF_1}$ and
$(S_{\CF_0})_{\CF_d} =S_{\CF_1}$ hold for coproducts and antipodes
(see Appendix for the proof).

The main advantage to decompose a twist element is that we can use
the isomorphism $\CA_{\CF_0}\simeq \hat{\CA}_0$~(\ref{diagram}).
Then, the second twist by $\CF_d$ in~(\ref{2step twist})
can also be regarded as a twist of the normal ordered module
algebra (vertex operators).
This enables us to study the ef\/fect of a~change of background
in terms of vertex operators, as we will see.

First, the twisted product of two local functional $F,G \in \CA_{\CF_1}$ is rewritten
by using (\ref{F_1=F_BF_0}) as
\begin{gather}
F*_{\CF_1} G
 =  m \circ \CF_1^{-1} \act (F\otimes G)
 =  m \circ \Delta\big(\CN_0^{-1}\big)(\CN_0 \otimes \CN_0) \CF_d^{-1} \act (F\otimes G) \nonumber\\
\phantom{F*_{\CF_1} G}{}
= \CN_0^{-1} \act m \circ \tilde{\CF_d}^{-1}
(\CN_0 \act F \otimes \CN_0 \act G)
 =  \CN_0^{-1} \act \big(:\!F\!: *_{\tilde{\CF_d}} :\!G\!: \big),
\label{twisted normal product}
\end{gather}
where the coboundary relation (\ref{coboundaryrelation}) for $\CF_0^{-1}$ is used.
Note that the dressed version of the twist element, def\/ined by
$\tilde{\CF_d}=(\CN_0 \otimes \CN_0)\CF_d (\CN_0^{-1} \otimes \CN_0^{-1})
\in \hat{\CH}_0 \otimes \hat{\CH}_0$ appears in the expression,
because it should act on $\hat{\CA}_0 \otimes \hat{\CA}_0$.
In fact, it is shown that $\tilde{\CF_d}$ is a twist element in $\hat{\CH}_0$.
In the same manner, the action of $h \in \CH_{\CF_1}$ on the same product
is written as
\begin{gather}
h  \act \big( F*_{\CF_1} G \big)
 = \CN_0^{-1} \tilde{h}  \act \big(:\!F\!: *_{\tilde{\CF_d}} :\!G\!: \big)
 = \CN_0^{-1} \act m \circ {\tilde{\CF_d}}^{-1} \Delta_{\tilde{\CF_d}} (\tilde{h})
\act \big(:\!F\!: \otimes :\!G\!: \big),
\label{twisted normal action}
\end{gather}
where again $\tilde{h}=\CN_0 h \CN_0^{-1}$ acts on $\hat{\CA}_0$.
These relations (\ref{twisted normal product}) and (\ref{twisted normal action})
show that $\CA_{\CF_1}$ is isomorphic to the twisted module algebra
$({\hat{\CA}}_0)_{\tilde{\CF_d}}$ under the map $F\to :\!F\!:$ by the $\CN_0$ action.
The structure is summarized as
\begin{gather}
\label{twisted normal str}
\xymatrix{
\CA\,\, \ar[rr]^{\text{twist by}\,\CF_0\,\,} & {} &
\,\,\CA_{\CF_0}\,\, \ar[d]_{\CN_0 \act} \ar[rr]^{\text{twist by}\,\CF_d\,}
&& \,\,\CA_{\CF_1}\,\, \ar[d]^{\CN_0 \act} \\
& & {\hat{\CA}_0}\,\, \ar[rr]^{\text{twist by}\,\tilde{\CF_d}\,}
&& \,\,({\hat{\CA}}_0)_{\tilde{\CF_d}}
}
\end{gather}
The VEV for $\CA_{\CF_1}$ is also written in terms of
$({\hat{\CA}}_0)_{\tilde{\CF_d}}$ as
\begin{gather*}
\tau (F*_{\CF_1} G)
 = \tau \circ \CN_0^{-1} \act \big(:\!F\!: *_{\tilde{\CF_d}} :\!G\!: \big)
 = \langle  :\!F\!:*_{\tilde{\CF_d}} :\!G\!:  \rangle_0.
\label{twisted normal VEV}
\end{gather*}
Thus the ef\/fect of the second twisting by $\CF_d$ is completely incorporated
as the twisting of the normal ordered Hopf algebra ${\hat{\CH}}_0$
and module algebra ${\hat{\CA}}_0$.
It means that under the decomposition~(\ref{F_1=F_BF_0}), the f\/irst
twist $\CF_0$ serves as a quantization w.r.t.\ a f\/ixed background,
and the second twist~$\CF_d$ play the role of the deformation of the
vertex operator product with keeping the background.

\subsection{Relating two backgrounds}

The above procedure is possible for any $\CF_d$ if it exists.
Let us now specialize the situation, in which
$\CF_1$ has also the form~(\ref{eq:F}) for some propagator $G_1$.
Then it also def\/ines a quantization scheme, which is achieved by the twist element
as in~(\ref{eq:F})
\begin{gather}
\CF_1:=\exp\left\{-\int   d^2z   \int   d^2w \,
G_1^{\mu\nu}(z,w)
\frac{\delta}{\delta X^\mu(z)}
\otimes \frac{\delta}{\delta X^\nu(w)}\right\} .
\label{F_1}
\end{gather}
Namely, we obtain the twisted pair $(\CH_{\CF_1},\CA_{\CF_1})$
as well as the normal ordered pair $(\hat{\CH}_1,\hat{\CA}_1)$
of Hopf and module algebras as in Section~\ref{sec:twist quantization}.
We denote this normal ordering
as $ ^{\circ}_{\circ} \cdots  ^{\circ}_{\circ} $, then for example
\begin{gather*}
 {}^{\circ}_{\circ} F*_{\CF_1}G{}^{\circ}_{\circ}
= {}^{\circ}_{\circ} F {}^{\circ}_{\circ} {}^{\circ}_{\circ} G {}^{\circ}_{\circ}
\end{gather*}
holds.

From the assumption that both $\CF_0$ and $\CF_1$ has the form (\ref{eq:F}),
the deviation $\CF_d$ in (\ref{F_1=F_BF_0}) has necessarily the form (\ref{eq:F}).
It is equivalent to the decomposition of the propagator
\begin{gather*}
G_1^{\mu\nu} (z,w)=G_0^{\mu\nu} (z,w)+G_d^{\mu\nu} (z,w),
\end{gather*}
which is the situation discussed in \cite {AMW2}.
It is now possible to relate $\hat{\CA}_1$ and $({\hat{\CA}}_0)_{\tilde{\CF_d}}$,
and this gives an additional structure to (\ref{twisted normal str})
as follows.
First, note that $\tilde{\CF_d}=\CF_d$ in this case.
We denote two normal ordered functionals as $:\!F\!: \in \hat{\CA}_0$
and ${}^{\circ}_{\circ} F{}^{\circ}_{\circ} \in \hat{\CA}_1$.
According to (\ref{F_1=F_BF_0}), two normal orderings are also related as
$\CN_1=\CN_d \CN_0$.
Then, we have a relation
${}^{\circ}_{\circ} F {}^{\circ}_{\circ} = \CN_d  \act :\!F\!:$ between them.
The relation is such that the VEVs, each with respect to the corresponding
quantization scheme, give the same result
\begin{gather*}
\langle{}^{\circ}_{\circ} F {}^{\circ}_{\circ} \rangle_1
= \tau \left( \CN_1^{-1} \act {}^{\circ}_{\circ} F {}^{\circ}_{\circ} \right)
= \tau \left( \CN_0^{-1} \act :\! F \!:\right)
= \langle  :\!F\!:  \rangle_0.
\end{gather*}
The dif\/ference of two quantization schemes lies in the Wick contraction
of several local functionals.
In fact, applying $\tau$ on both sides of (\ref{twisted normal product})  leads to the equation of the VEV
\begin{gather*}
\langle{}^{\circ}_{\circ} F {}^{\circ}_{\circ} {}^{\circ}_{\circ} G
{}^{\circ}_{\circ}  \rangle_1
 = \tau (F*_{\CF_1} G)
 = \langle  :\!F\!:*_{\CF_d} :\!G\!:  \rangle_0.
\end{gather*}
On the l.h.s., the def\/inition of the VEV as well as the normal
ordering are
with respect to the new quantization $\CF_1$.
The r.h.s.\ is written in terms of the original quantization scheme~$\CF_0$
except that the OPE is twisted by $\CF_d$,
that is $(\hat{\CA}_0)_{\CF_d}$ in~(\ref{twisted normal str}).
Similarly, the twisted Hopf algebra $(\hat{\CH}_0)_{\CF_d}$ is
mapped to the normal ordered Hopf algebra $\hat{\CH}_1$ by
$\tilde{h} \to \CN_d \tilde{h} \CN_d^{-1}=\CN_1 h \CN_1^{-1}\equiv {\tilde{h}}_1$
and we have
\begin{gather*}
\langle \tilde{h}_1 \act {}^{\circ}_{\circ} F {}^{\circ}_{\circ} {}^{\circ}_{\circ} G
{} ^{\circ}_{\circ}  \rangle_1
 = \langle  \tilde{h}  \act \left(:\!F\!: *_{\CF_d} :\!G\!: \right)  \rangle_0,
\end{gather*}
Let us summarize the whole structure:
\begin{gather}
\xymatrix{
\CA\,\, \ar[rr]^{\text{twist by}\,\CF_0\,\,} & {} &
\,\,\CA_{\CF_0}\,\, \ar[d]_{\CN_0 \act} \ar[rr]^{\text{twist by}\,\CF_d\,}
&& \,\,\CA_{\CF_1}\,\, \ar[d]^{\CN_0 \act} \\
& & {\hat{\CA}_0}\,\, \ar[rr]^{\text{twist by}\,\CF_d\,}
&& \,\,({\hat{\CA}}_0)_{\CF_d} \ar[d]^{\CN_d \act}\\
&&&& \hat{\CA}_1
}
\label{triangle maps}
\end{gather}
In this way, given two backgrounds $\CF_0$ and $\CF_1$,
the corresponding quantum operator algebras and symmetries
are related through the deformation $\CF_d$, and since
 $\CF_1=\CF_d \CF_0=\CF_0 \CF_d$,
we can also exchange the role of $\CF_0$ and $\CF_d$.

\section{Applications}\label{section5}

Here we apply the method given in Section~\ref{sec:relating} to three cases:
the constant $B$-f\/ield background, target space dif\/feomorphism,
and extra twist in zero modes.
The f\/irst is already appeared in~\cite{AMW2} and the others are new.

\subsection[Twist quantization with $B$ field]{Twist quantization with $\boldsymbol{B}$ f\/ield}
\label{sec:H_1}

As an application of the above mentioned method,
let us consider the background with a constant $B$-f\/ield.
For simplicity, we take the worldsheet $\Sigma$ to be the upper half plane
of the complex plane.
Then, the propagator satisfying the mixed boundary condition is given as
\cite{Fradkin:1985qd,1987NuPhB.288..525C,1987NuPhB.280..599A}
\begin{gather}
G_1^{\mu\nu}(z,w)  = -{\alpha'}  \left[
\eta^{\mu\nu}\ln |z-w| -\eta^{\mu\nu}\ln |z-\bar w|
+G^{\mu\nu}\ln |z-{\bar w}|^2 +  {\Theta^{\mu\nu}} \ln
\frac{z-{\bar w}}{{\bar z}-w} \right],
\label{UHPpropagator}
\end{gather}
where the open string metric $G^{\mu\nu}$ and the antisymmetric tensor
$\Theta^{\mu\nu}$ are def\/ined by
\begin{gather*}
G^{\mu\nu}
=\left(\frac{1}{\eta+B} \eta \frac{1}{\eta-B} \right)^{\mu\nu}
\qquad \mbox{and}\qquad  \Theta^{\mu\nu}= \frac{\theta^{\mu\nu}}{2\pi \alpha'}
= -\left(\frac{1}{\eta+B} B \frac{1}{\eta-B} \right)^{\mu\nu}.
\label{G_1}
\end{gather*}

We take the twist $\CF_1$ of the form~(\ref{eq:F}) with this propagator.
Twisting by $\CF_1$ directly, we obtain the normal ordered module algebra $\hat{\CA}_1$.
The same normal ordering in the context of operator formalism is given in the literature
\cite{Braga:2004wr,Braga:2006gi,Chakraborty:2006yj,Gangopadhyay:2007ne}.
Here we decompose $\CF_1$ in two dif\/ferent ways \cite{AMW2}.

\subsubsection{Relation to f\/ield theory and twisted Poincar\'e symmetry}
\label{sec:field theory}

The f\/irst decomposition is meaningful only
for the open string case.
By restricting the functional space to the
one corresponding to the open string vertex operators,
we can derive the relation to the
twisted Poincar\'e symmetry in the f\/ield theory on the noncommutative
space considered in \cite{Chaichian:2004za,Koch:2004ud, Aschieri:2005yw}.

To this end, we decompose the propagator into
the symmetric part $G_S$ and the anti-symmetric part $G_A$ with respect to
the tensor indices $\mu \nu$:
\begin{gather*}
G_1^{\mu\nu}(z,w)=G_S^{\mu\nu}(z,w)+G_A^{\mu\nu}(z,w).
\end{gather*}
At the boundary $s,t \in \p\Sigma=\real$, they reduce to the form
\begin{gather*}
G_S^{\mu\nu}(s,t)=-{\alpha'}G^{\mu\nu}\ln(s-t)^2
\qquad \mbox{and}\qquad G_A^{\mu\nu}(s,t)=\frac{i}{2} {\theta^{\mu\nu}} \epsilon (s-t),
\label{SA boundary}
\end{gather*}
where $\epsilon (t) $ is the sign function.
This implies the decomposition $\CF_1=\CF_A \CF_S$ (\ref{F_1=F_BF_0}).
Under the f\/irst twist,
we have a twisted module algebra $\CA_{\CF_S}$ with the
product $*_{\CF_S}$, and the
normal ordered module algebra $\hat{\CA}_S$ (see (\ref{triangle maps})).
We denote elements of $\hat{\CA}_S$ as
$\,^{\bullet}_{\bullet} F\,^{\bullet}_{\bullet} =\CN_S \act F$.
Clearly, this def\/ines a quantization scheme w.r.t.\ the open string metric
as a background, but it turns out that
it is natural only for boundary elements of $\CH$ and $\CA$.\footnote{They are def\/ined by the functionals and functional derivatives
of the embedding $X^\mu (t)$ of the boundary, or equivalently, def\/ined by inserting
a delta function of the form $\int_{\p\Sigma} dt\delta^{(2)} (z-t)$ into $\CH$ and $\CA$.}

The quantum (untwisted) symmetry in this case is easily found.
Although the Lorentz generator $L_{\mu\nu} \in \CP$ in (\ref{generators})
acquire the twist and become non-primitive,
there are other boundary elements $L'_{\mu\nu} \in \CH$ of the form
\begin{gather*}
L'_{\mu\nu} = \int_{\p\Sigma}   dt \,
G_{[\mu \rho}X^{\rho}(t) \frac{\delta}{\delta X^{\nu]}(t)}
\label{L'}
\end{gather*}
such that $L'_{\mu\nu}$ and the translation generators $P_{\mu}$
(as boundary elements) constitute
another Poincar\'e--Lie algebra $\CP'$ when acting on boundary local functionals,
where the commutation relations are written
with respect to the open string metric $G_{\mu\nu}$.
It is easy to show that the Hopf subalgebra $U(\CP')$ of $\CH$ is invariant
under the twist $\CF_S$.
In terms of the normal ordered Hopf algebra,
this means
$\tilde{P_{\mu}}=P_{\mu}$ and $\tilde{L}'_{\mu\nu}=L'_{\mu\nu}$
as boundary elements in $\hat{\CH}_S$.
Therefore,
$U(\CP')$ is considered to be a quantum Poincar\'e symmetry in this quantization scheme
when restricted on the boundary.

As a consequence,
boundary elements of the module algebra $\hat{\CA}_S$ are classif\/ied by
the representation of $U(\CP')$ with a f\/ixed momentum~$k_\mu$.
In general, a local boundary vertex opera\-tor~$V_k (t)$
with momentum $k_\mu$
 consists of the worldsheet derivatives of $X$ and $e^{ik\cdot X}$,
having the form
\begin{gather}
V_k (t)={}^{\bullet}_{\bullet} P[\p X(t)] e^{ik\cdot X(t)} {}^{\bullet}_{\bullet} .
\label{boundary vertex}
\end{gather}
Here $P[\p X]$ denotes a polynomial.
Note these vertex operators $V_k (t)$ are
equivalent to the vertex operators used in~\cite{Seiberg:1999vs},
because their anomalous dimension and the on-shell condition are determined
with respect to the open string metric $G_{\mu\nu}$.

Twisting by $\CF_A$ further,
we obtain  the twisted Hopf algebra $(\hat{\CH}_S)_{\CF_A}$
and the twisted module algebra $\CA_{\CF_1} \simeq (\hat{\CA}_S)_{\CF_A}$
(see (\ref{triangle maps})).
The twisted product $*_{\CF_1}$ in $\CA_{\CF_1}$
is seen as a deformation of the product in $\hat{\CA}_S$ as above:
\begin{gather*}
\langle{}^{\circ}_{\circ} F {}^{\circ}_{\circ} {}^{\circ}_{\circ} G {}^{\circ}_{\circ}
 \rangle_1
= \tau( F*_{\CF_1} G )
= \langle {}^{\bullet}_{\bullet} F{}^{\bullet}_{\bullet}
*_{\CF_A} {}^{\bullet}_{\bullet}G{}^{\bullet}_{\bullet} \rangle_S.
\end{gather*}
By considering only the boundary vertex operators,
the twist element acting on them has the form
\begin{gather}
\CF_A  =  \exp\left\{-\frac{i}{2} {\theta^{\mu\nu}}
\int   ds   \int   dt \,\epsilon (s-t)
\frac{\delta}{\delta X^\mu(s)}
\otimes \frac{\delta}{\delta X^\nu(t)}\right\} .
\label{F_A}
\end{gather}
Because worldsheet derivatives of $X$ do not feel this deformation:
e.g.,
$\p_a X^{\mu}(s) *_{\CF_A} X^\nu(t)= \p_a X^{\mu}(s) X^\nu(t)$,
for any correlation function of boundary vertex operators of the form
(\ref{boundary vertex}),
the product $*_{\CF_A}$ is only sensitive to the exponential part, and
we have
\begin{gather*}
\langle V_{k_1}(t_1) *_{\CF_A} \cdots *_{\CF_A} V_{k_n}(t_n)  \rangle_S
= e^{-\frac{i}{2}\sum_{i>j} k_{i\mu} \theta^{\mu\nu} k_{j\nu}
\epsilon (t_i-t_j) }
\langle V_{k_1}(t_1) \cdots V_{k_n}(t_n)  \rangle_S .
\label{correlation function}
\end{gather*}
In order to move from correlation functions on the worldsheet
to the ef\/fective theory on D-branes, we should f\/ix the cyclic ordering
of the insertion points $t_1 > t_2 > \cdots > t_n$ \cite{Seiberg:1999vs}.
Then the extra phase factor
in the r.h.s.\ above becomes independent of the precise locations,
and gives the factor of the Moyal product.
This enables us to take the prescription~\cite{Seiberg:1999vs}:
{\it replace ordinary multiplication in the ef\/fective f\/ield theory written in the
open string metric by the Moyal product}.
In the Hopf algebra language, this means that
with a f\/ixed ordering the twist element (\ref{F_A}) acts as
a Moyal-twist \cite{Oeckl:2000eg, Watts:2000mq}
\begin{gather*}
\CF_M = e^{\frac{i}{2} {\theta^{\mu\nu}} P_\mu \otimes P_\nu} ,
\end{gather*}
on f\/ields on the D-brane.
Here, the integrals of functional derivatives in~(\ref{F_A})
are replaced by its zero mode part, the translation generators $P_\mu \in U(\CP')$.

The Poincar\'e symmetry on D-branes is also read of\/f under the above prescription.
The coproduct of the Lorentz generators $L'_{\mu\nu} \in U(\CP')$
acting on boundary vertex operators is twisted by $\CF_A$ as
\begin{gather*}
\Delta_{\CF_A} (L'_{\mu\nu}) = \Delta (L'_{\mu\nu})
+\frac{1}{2} {\theta^{\alpha \beta }}
\int  ds   \int   dt \,\epsilon (s-t) G_{\alpha[\mu }
\frac{\delta}{\delta X^{\nu]}(s)}
\otimes \frac{\delta}{\delta X^\beta (t)}
\end{gather*}
but it reduces by f\/ixing the ordering $s>t$ to
\begin{gather*}
\Delta_{\CF_M} (L'_{\mu\nu}) =\Delta (L'_{\mu\nu})
+\frac{1}{2} {\theta^{\alpha \beta }}
\left\{
 G_{\alpha[\mu }P_{\nu]} \otimes P_\beta
 + P_\alpha  \otimes G_{\beta [\mu } P_{\nu]}
\right\}.
\end{gather*}
This is nothing but the twisted Poincar\'e--Hopf algebra
$U_{\CF_M} (\CP')$ found in \cite{Chaichian:2004za,Koch:2004ud,Aschieri:2005yw}.
Note that since the twist element
itself now belongs to $U(\CP') \otimes U(\CP')$,
the twisting is closed in $U(\CP')$.

To summarize, the f\/irst twist is a quantization
with respect to the open string metric, when restricted to boundary vertex operators,
and the second is its deformation, which reduces to the Moyal-twist on the D-brane
worldvolume.
In other words, the twisted Poincar\'e symmetry on the Moyal--Weyl noncommutative space
is derived from string worldsheet theory in a $B$-f\/ield background.
Thus, we establish the diagram schematically
\begin{gather*}
\begin{array}{@{}ccc}
(\hat{\CH}_S)_{\CF_A} & \xrightarrow{\text{f\/ield theory}} & U_{\CF_M}(\CP') \\
\triangledown && \triangledown \\
(\hat{\CA}_S)_{\CF_A} & \xrightarrow{\text{f\/ield theory}} & A_{\CF_M}.
\end{array}
\label{string-field diagram}
\end{gather*}

\subsubsection[$B$ field as a deformation]{$\boldsymbol{B}$ f\/ield as a deformation}
\label{sec:B as deform}

The next decomposition is more natural from the viewpoint of
the twist quantization.
There we can see the relation
between the standard quantization and the deformation
caused by a $B$-f\/ield background.
With this decomposition, we obtain a new twisted Poincar\'e symmetry,
which is dif\/ferent from \cite{Chaichian:2004za,Koch:2004ud, Aschieri:2005yw}
realized only on D-branes.

Here we consider a decomposition
of the propagator (\ref{UHPpropagator}) into the $B=0$ part and $B$-dependent part:
\begin{gather*}
  G_1^{\mu\nu}(z,w) = G_0^{\mu\nu}(z,w) + G_B^{\mu\nu}(z,w) ,
\label{G_1=G_0+G_B}\\
  G_B^{\mu\nu}(z,w)=-{\alpha'} \left[
(G-\eta)^{\mu\nu} \ln |z-{\bar w}|^2 +  {\Theta^{\mu\nu}} \ln
\frac{z-{\bar w}}{{\bar z}-w}\right] ,
\label{propagatorwithB}
\end{gather*}
where $G_0^{\mu\nu}(z,w)$ is given in (\ref{Cpropagator}).
Accordingly, the twist element $\CF_1$ (\ref{F_1}) is divided into
\begin{gather*}
\CF_1=\CF_B \CF_0 .
\label{B decomposition}
\end{gather*}
Here the ef\/fect of the $B$-f\/ield is completely characterized
as a (second) twist $\CF_B$ of the Hopf algebra $\CH_{\CF_0}$
and the module algebra $\CA_{\CF_0}$,
that are regarded as a deformation of $\hat{\CH}_0$
and the vertex operator algebra $\hat{\CA}_0$:
\begin{gather*}
\langle{}^{\circ}_{\circ} F {}^{\circ}_{\circ} {}^{\circ}_{\circ} G
{}^{\circ}_{\circ}  \rangle_1
 = \langle  :\!F\!:*_{\CF_B} :\!G\!:  \rangle_0.
\end{gather*}

Let us now focus on the fate of the Poincar\'e symmetry $U(\CP)$.
Under the f\/irst twist, $U(\CP)$ is a twist
invariant Hopf subalgebra of $\CH_{\CF_0}$.
Equivalently, $U(\CP)$ is a Hopf subalgebra of $\hat{\CH}_0$
with $\tilde{P}_{\mu}=P_{\mu}$ and $\tilde{L}_{\mu\nu}=L_{\mu\nu}$
as elements in $\hat{\CH}_0$.
Thus, $U(\CP)$ remains a symmetry at the quantum level, and each (normal ordered)
vertex operator in $\hat{\CA}_0$ is in a representation of a Poincar\'e--Lie algebra.
We emphasize that this fact guarantees
the spacetime meaning of a vertex operator.
For example, a graviton vertex operator
$V=:\!\p X^\mu \bar{\p}X^\nu e^{ikX}\!:$ transforms as spin $2$ representation
under $U(\CP)$, corresponding to the spacetime graviton f\/ield $h_{\mu\nu}(x)$.

However, in a $B$-f\/ield background,
the second twist $\hat{\CH}_0 \rightarrow (\hat{\CH}_0)_{\CF_B}$
modif\/ies the coproduct and antipode of the
Lorentz generator in
$U(\CP)$ as
\begin{gather}
 \Delta_{\CF_B} (L_{\mu\nu}) = L_{\mu\nu}\otimes 1+1\otimes L_{\mu\nu}
-2\int d^2 z d^2 w \, \eta_{[\mu\alpha} G_B^{\alpha\beta }(z,w)
\frac{\delta}{\delta X^\beta (z)}\otimes \frac{\delta}{\delta X^{\nu]} (z)},
\label{twisted coproduct for L}\\
 S_{\CF_B}(L_{\mu\nu}) =S(L^{\mu\nu})+
2\int d^2ud^2z\,G_B^{\rho[\mu}(u,z)\frac{\delta}{\delta X^{\rho}(u)}
\frac{\delta}{\delta X_{\nu]}(z)} .
\label{twisted antipode for L}
\end{gather}
Apparently, this $L_{\mu\nu}$ is not primitive and this means
that the $U(\CP)$ is twisted.
We give several remarks about this twisted Poincar\'e symmetry read
from (\ref{twisted coproduct for L}), (\ref{twisted antipode for L}):
\begin{enumerate}\itemsep=0pt
\item
The twisting of $U(\CP)$ is only due to the second twist $\CF_B$.
This guarantees that a single local vertex operator $:\!F\!:$
in $(\hat{\CA}_0)_{\CF_B}$ is still Poincar\'e covariant after the twist,
and thus the spacetime
meaning of a vertex operator is unchanged.
\item
The twisting is not closed within $U(\CP)$,
but the additional terms in (\ref{twisted coproduct for L}) are
in $U({\mathfrak C})\otimes U({\mathfrak C})$,
and similarly for (\ref{twisted antipode for L}).
This shows that the twisting does not mix the Lorentz generator with other
dif\/feomorphisms even for products of vertex operators.
Therefore, from the spacetime point of view,
a Poincar\'e transformation on products of f\/ields
is still a global transformation even after the twist.
\item The twisted product $*_{\CF_B}$ of two vertex operators
reduces to the product of corresponding spacetime f\/ields,
but the latter depends on the representations (spins) of $U(\CP)$,
because $\CF_B$ depends on the Green function $G_B$.
For instance, the twisted product of two gravitons is dif\/ferent from
that of two tachyons due to this fact.
Similarly, the Lorentz transformation law
for the product of two spacetime
f\/ields depends on their spins.
\end{enumerate}

To summarize, from the decomposition of the twist element $\CF_1=\CF_B \CF_0$,
we obtain a description of a spacetime with a $B$-f\/ield background,
such that the ef\/fect of background $B$-f\/ield is
hidden in the twist element $\CF_B$,
while the other matter f\/ields acquires a deformation
caused by this twist element.
This description of spacetime is valid for closed strings as well,
and is dif\/ferent from the description formulated on
a commutative spacetime with a matter f\/ield $B_{\mu\nu}$ or
from the description using
a noncommutative space as in the ef\/fective theory of open string.
Although it is not clear yet how this new description translates
into the ef\/fective f\/ield theory language, our method clearly demonstrates
how a background f\/ield
(other than the metric) can be incorporated into a formulation of a stringy spacetime.

\subsection{Twisted dif\/feomorphism}
\label{sec:finite diffeo}

Here we discuss about the twisted dif\/feomorphism
in the context of successive twists.
For simplicity, we f\/ix a twist element $\CF_0$ to be Minkowski background,
but the following argument can be applied to any f\/ixed $\CF_0$
of the form (\ref{eq:F}).

Let $\xi \in {\mathfrak X}$ be a functional vector f\/ield of some target space vector f\/ield.
Its f\/inite version $u=e^{\xi} \in \CH=U({\mathfrak X})$ is group-like
$\Delta (u)=u\otimes u$.
It acts on the coordinate functional as
$u \act X^\mu (z)=X^\mu (z)+\xi^\mu [X(z)] +{\cal O}(\xi^2)\equiv Y^\mu (z)$.
Thus the action of $u$ is interpreted as a~coordinate
transformation (from the passive viewpoint).
Classically, it is a automorphism of the algebra of functionals
$u\act (F\cdot G)=F_u \cdot G_u$,
where $F_u [X]=(u\act F)[X]=F[Y]$ is understood
as a new functional written in the transformed coordinates.
However, this is not the case for $\CA_{\CF_0}$,
that is seen by rewriting the action of $u$ on the twisted product
\begin{gather*}
u\act (F*_{\CF_0}G)
 =  m \circ \CF_0^{-1} \Delta_{\CF_0}(u) \act (F\otimes G)
 =  m \CF_1^{-1} \act (u\act F\otimes u\act G)
 =  F_u *_{\CF_1} G_u.
\end{gather*}
Namely, from the passive viewpoint, the product is also needed to be transformed
to give a new twist element \cite{AMW}
\begin{gather}
\CF_1
= (u\otimes u)\CF_0 \Delta \big(u^{-1}\big).
\label{group coboundary}
\end{gather}
Note that (\ref{group coboundary}) is also a coboundary twist
since it can be written in the form (\ref{coboundaryrelation}) with
the new normal ordering $\CN_1=u\CN_0 u^{-1}$.
As a consequence, there are two vertex operator algebras~$\hat{\CA}_0$ and~$\hat{\CA}_1$ and they are related by the $u$-action as
${}^{\circ}_{\circ} F_u {}^{\circ}_{\circ} = \CN_1 u \act F = u \CN_0 \act F
=u \, \act :\!F\!:$.
It is natural that the vertex operator algebra transforms
under the dif\/feomorphism, because the change of the background needs a
new mode expansion in the operator formulation.
However, $u \, \act :\!F\!:$ is not well-def\/ined as operators since
$u$ is not an element in~$\hat{\CH}_0$.
In other words, $u \, \act :\!F\!:$ contains formal divergent terms.
See \cite{AMW} for more details to come to this point.
Here we would like to discuss that this  dif\/f\/iculty is solved by applying
the method in Section~\ref{sec:relating}.

To this end, we will f\/irst decompose $\CF_1$ (\ref{group coboundary})
to the form $\CF_1=\CF_d \CF_0$.
It is easy to show that
\begin{gather}
\CF_d = (u\otimes u)\Delta_{\CF_0 } \big(u^{-1}\big)
\label{Fd}
\end{gather}
is the desired twist element in $\CH_{\CF_0}$.
Note that~(\ref{Fd}) is not in $U({\mathfrak C})\otimes U({\mathfrak C})$ in general,
nor of the type~(\ref{eq:F}).
By applying the rule~(\ref{twisted normal str}),
the deformation of the vertex operator algebra $\hat{\CA}_0$
to~$(\hat{\CA}_0)_{\tilde{\CF_d}}$ is governed by
\begin{gather*}
\tilde{\CF_d} =(\CN_0\otimes \CN_0) \CF_d \big(\CN_0^{-1}\otimes \CN_0^{-1}\big)
 =(\CN_0\otimes \CN_0)(u\otimes u)\CF_0 \big(u^{-1}\otimes u^{-1}\big) \CF_0^{-1}
\big(\CN_0^{-1}\otimes \CN_0^{-1}\big) \nonumber\\
\phantom{\tilde{\CF_d}}{}
 =(\CN_0\otimes \CN_0)(u\otimes u)\big(\CN_0^{-1}\otimes \CN_0^{-1}\big) \Delta (\CN_0)
\Delta (u^{-1}) \Delta \big(\CN_0^{-1}\big)
 =(\tilde{u}\otimes \tilde{u}) \Delta \big(\tilde{u}^{-1}\big).
\end{gather*}
This does not reduce to $1$ because $\tilde{u} =\CN_0 u \CN_0^{-1} \in \hat{\CH}_0$
is not group-like.
Thus the twisting by~$\tilde{\CF_d}$ must deform the operator product.
In $(\hat{\CA}_0)_{\tilde{\CF_d}}$, a single vertex operator and a product of two
vertex operators are related to that of $\hat{\CA}_0$ as
$:\!F_u \!: =\tilde{u}~ \act :\!F\!:$ and
\begin{gather*}
:\!F_u \!:*_{\tilde{\CF_d}} :\!G_u \!:
 =m \circ \tilde{\CF_d}^{-1} \act \left( :\!F_u \!: \otimes :\!G_u \!: \right)
 =m \circ \Delta (\tilde{u}) \big(\tilde{u}^{-1} \otimes \tilde{u}^{-1}\big)
\act \left( :\!F_u \!: \otimes :\!G_u\!: \right) \nonumber\\
\phantom{:\!F_u \!:*_{\tilde{\CF_d}} :\!G_u \!:}{}
 =\tilde{u} \act \left( :\!F\!:\, :\!G\!: \right) ,
\end{gather*}
which are given by well-def\/ined actions of $\tilde{u}$.

We conclude that any background dif\/feomorphic to the original
background $\CF_0$ is also described by a twist~$\CF_1$~(\ref{group coboundary}).
In the vertex operator language, this causes a deformation of the origi\-nal
OPE, without changing the background itself.
We can now see the ef\/fect of dif\/feomorphism by comparing~$\hat{\CA}_0$ and
$(\hat{\CA}_0)_{\tilde{\CF}_d}$ instead of $\hat{\CA}_0$ and $\hat{\CA}_1$.

\subsection{Zero modes}
\label{sec:zero mode}

In this section, we discuss some results about the role of the
zero modes in twist quantization.
Let us f\/ix a twist element $\CF_0$ for the Minkowski background,
which is given by the Green func\-tion~(\ref{Cpropagator}).
Our convention on the VEV def\/ined in
(\ref{def of true VEV}) for $\hat{\CA}_0$
has the properties
\begin{gather}
\langle  1  \rangle_0 =1, \qquad
\langle  X^\mu (z)  \rangle_0 =0, \qquad
\langle  :\! e^{ik\cdot X(z)}\!:  \rangle_0 =1.
\label{original convention}
\end{gather}
These do not match with the conventional correlation functions,
that have momentum conservation in target spacetime\footnote{They are conventional correlation functions as a two dimensional scalar f\/ield theory.}.
Here we argue that this dif\/f\/iculty can be removed easily by changing the
constant term in the propagator, or equivalently, a further twisting
on the zero modes.

Motivated by a tachyon one point function satisfying the momentum conservation
\begin{gather*}
\langle  :\!e^{ik \cdot X(z)}\!:  \rangle \propto (2\pi)^d \delta^{(d)}(k)  ,
\end{gather*}
we def\/ine a map $\tilde{\tau}: \CA \to \complex$ by
\begin{gather}
\tilde{\tau} (F[X]) := \int   d^d x \,F[X]|_{X=x} ,
\label{new def of VEV}
\end{gather}
that is, we set $X^\mu (z)=x^\mu$ in the functional
instead of setting $X=0$ as in
(\ref{def of true VEV}) and then integrate over~$x$.
This is a consistent treatment comparing with the path integral quantization.
There, since the zero modes~$x^\mu$ of $X^\mu(z)$ do not appear
in the action, they are integrated out separately from non-zero modes.
(\ref{new def of VEV}) corresponds to this integration.
Note that the integrand in (\ref{new def of VEV}) is related to $\tau$ as
$F[X]|_{X=x} = \tau (F[X+x]) = \tau (e^{-ix^\mu P_\mu} \act F[X])$,
where $P_\mu \in U({\mathfrak C})$ is the translation generator in~(\ref{generators}).
If we use a following relation as an operator\footnote{The Wick rotation should be performed in the time direction.}
\begin{gather*}
\int   d^d x \,e^{-ix^\mu P_\mu} = (2\pi)^d \delta^{(d)}(P)
= (2\pi)^d  \lim_{C\to \infty} \left(\frac{C}{2\pi}\right)^{\frac{d}{2}}
e^{-\frac{C}{2}\eta^{\mu\nu} P_\mu P_\nu},
\end{gather*}
then the map $\tilde{\tau}$ is also written as
\begin{gather*}
\tilde{\tau}\left(F[X] \right)=\lim_{C\to\infty}(2\pi C)^{\frac{d}{2}}
\tau\left(\CN_d^{-1} \act F[X] \right) ,
\end{gather*}
where we def\/ined a Hopf algebra element $\CN_d \in U({\mathfrak C})$ by
\begin{gather}
\CN_d=e^{\frac{C}{2}\eta^{\mu\nu} P_\mu P_\nu}.
\label{def of N_d}
\end{gather}
In the following,
$C$ is always understood as a positive divergent constant.
For example,
\begin{gather}
\tilde{\tau}(1) = (2\pi C)^{d/2}, \qquad
\tilde{\tau}\big(e^{ik\cdot X(z)}\big) = (2\pi)^d \delta^{(d)}(k).
\label{ex1}
\end{gather}
By renormalizing the overall factor, we def\/ine a new map (with suf\/f\/ix $d$) as
\begin{gather}
\tau_d \left(F[X] \right)= \tau\left(\CN_d^{-1} \act F[X] \right).
\label{def of tau_d}
\end{gather}
For example,
\begin{gather*}
\tau_d (1)=1, \qquad
\tau_d \big(e^{ik\cdot X(z)}\big)
=\frac{(2\pi )^d \delta^{(d)}(k)}{(2\pi C)^{d/2}} .
\label{ex2}
\end{gather*}
The divergent overall factor $(2\pi C)^{d/2}$ in (\ref{ex1})
comes form the inf\/inite volume $\sim \delta^d (0)$ in the target space.
Thus the relation between the two maps $\tilde{\tau}$ and $\tau_d$
is exactly the same as that between the unnormalized
and normalized path integrals.

This new def\/inition of the VEV (\ref{def of tau_d})
suggests a relation to a new twisting.
We def\/ine $\CF_1$ of the form (\ref{eq:F})
with $G_1$ given by adding a (divergent) constant term $C$ as
\begin{gather*}
G_1^{\mu\nu}(z,w) = G_0^{\mu\nu}(z,w) + C\eta^{\mu\nu}.
\end{gather*}
This modif\/ication is, in fact, needed for the compact worldsheet~\cite{Polchinski}.
When calculating the Feynman propagator on the worldsheet,
such a divergent term in the Green function appears
as an IR cut of\/f, sometimes written as
$G_1^{\mu\nu}(z,w)=-\alpha' \eta^{\mu\nu}\ln \mu |z-w|$  $(0< \mu)$, thus $C=-\alpha'\ln \mu$.
From the target space point of view, the term ref\/lects the
non-compactness of the Minkowski spacetime
(i.e., zero mode part $\Bra{0}\hat{x}^\mu \hat{x}^\nu \Ket{0}$ in the two point function).

We can now apply the decomposition method to the twist
$\CF_1=\CF_d \CF_0$ and the normal ordering $\CN_1=\CN_d \CN_0$,
where $\CN_d$ is def\/ined in (\ref{def of N_d}) and
\begin{gather*}
\CF_d =e^{C\eta^{\mu\nu} P_\mu \otimes P_\nu},
\end{gather*}
but in the present case it is more appropriate to take the f\/irst twist by $\CF_d$,
and then to perform the second twisting by $\CF_0$.
This leads to a structure similar to (\ref{triangle maps}) as
\begin{gather}
\xymatrix{
\CA\,\, \ar[rr]^{\text{twist by}\,\CF_d\,\,} & {} &
\,\,\CA_{\CF_d}\,\, \ar[d]_{\CN_d \act} \ar[rr]^{\text{twist by}\,\CF_0\,}
&& \,\,\CA_{\CF_1}\,\, \ar[d]^{\CN_d \act} \\
& & \CA_d \,\, \ar[rr]^{\text{twist by}\,\CF_0\,}
&& \,\,(\CA_d)_{\CF_0} \ar[d]^{\CN_0 \act}\\
&&&& \hat{\CA}_1
}
\label{zero modes maps}
\end{gather}
In this diagram (\ref{zero modes maps}),
we denote the ``normal ordered" module algebra of the f\/irst twist $\CF_d$
as~$\CA_d$ (not as $\hat{\CA}_d$),
since the f\/irst twist by $\CF_d$ is not regarded as a quantization.
This is because
$\CN_d$ acts only on the zero mode part of functionals $F[X]$
and it does not cause any divergence coming from coincident points.
For example,
$\CN_d \act \p_a X^\mu (z) e^{ik \cdot X(z)}
= e^{\frac{C}{2} k^2} \p_a X^\mu (z) e^{ik \cdot X(z)}$
behaves as a~classical functional.
Correspondingly,
we prefer to write an element $\CA_d \ni F_d$ instead of $\CN_d \act F$.
The map $\CA_d \to \complex$ induced by $\tau: \CA \to \complex$
also does not have an interpretation as a VEV,
but it is the map $\tau_d$ given in~(\ref{def of tau_d}).
Indeed, we have a relation
$\tau_d (F_d)=\tau(\CN_d^{-1} \CN_d \act F)=\tau(F)$
as well as $\tau_d (F_d G_d)=\tau(FG)$.

Then, if we regard $\CA_d$ as a classical module algebra of functionals,
the second twist by~$\CF_0$ (a sub-diagram staring from $\CA_d$)
is interpreted as a quantization.
In particular, the VEV $(\CA_d)_{\CF_0} \to \complex$
is def\/ined like (\ref{def of true VEV}) but based on $\tau_d$ (not on $\tau$).
This is the main ef\/fect of the f\/irst twist $\CF_d$, which implies the momentum conservation.
The vertex operator algebra is now given by~$\hat{\CA}_1$,
whose elements have the form $:\! F_d \!:$, and the relation to $(\CA_d)_{\CF_0}$ is as usual:
\begin{gather*}
 \tau_d \left(F_d \right)
=\tau_d \left(\CN_0^{-1} :\! F_d \!: \right)
=\langle  :\! F_d \!:  \rangle_1 , \qquad
 \tau_d \left(F_d *_{\CF_0} G_d \right)
=\langle  :\! F_d \!: :\! G_d \!:  \rangle_1 .
\end{gather*}
Note the subscript 1 in the VEV $\langle  \cdots  \rangle_1$.
It depends both on $\CN_0^{-1}$ (the quantization) and
$\tau_d$ (the chosen vacuum state).
Of course, they equal to $\tau(F)$ and $\tau(F*_{\CF_1} G)$, respectively.

As an example, an ordinary tachyon $n$-point correlation function is given
by setting each local functional $F_d [X]=e^{ikX (z)}$ in $\CA_d$ as
\begin{gather*}
\langle  :\! e^{ik_1 X (z_1)} \!: \cdots :\! e^{ik_n X (z_n)} \!:  \rangle_1
 =\tau_d \Big( e^{ik_1 X (z_1)} *_{\CF_0} \cdots *_{\CF_0} e^{ik_n X (z_n)}\Big) \nonumber\\
 \qquad {} =\tau \Big( e^{-\frac{C}{2}k_1^2}e^{ik_1 X (z_1)} *_{\CF_1}
\cdots *_{\CF_1} e^{-\frac{C}{2}k_1^2}e^{ik_n X (z_n)}\Big) \nonumber\\
\qquad{} =e^{-\frac{C}{2}\sum_{i} k_i^2 -C\sum_{i<j} k_i \cdot k_j
+\alpha' \sum_{i<j} k_i \cdot k_j \ln |z_i-z_j|}
 = {\frac{(2\pi )^d \delta^{(d)}(\sum_{i} k_i)}{(2\pi C)^{d/2}}}
\prod_{i<j} |z_i-z_j|^{\alpha'  k_i \cdot k_j},
\end{gather*}
which is equivalent to starting with local functionals
$F[X]=e^{-\frac{C}{2}k^2}e^{ikX (z)}$ in $\CA$.
This example shows that all the constant $C$
appearing in the intermediate expression are cancelled among them
except for the overall normalization,
and the latter is also cancelled with the momentum delta-function
in the case of a momentum conserving correlation function.
Thus for renormalized functionals $F_d$,
it is designed so as to give a well-def\/ined vacuum expectation values.
On the other hand, this example also suggests
that a tachyon f\/ield $T(x)$ in spacetime corresponds to an already renormalized
functional $F_d[X]=e^{ikX(z)}$ on the worldsheet.
At this point, we have not yet fully understood the rule to give such a correspondence
between spacetime f\/ields and worldsheet functionals.
Since it needs both worldsheet and spacetime analysis, we leave it for a~further study.

In summary, the quantization with the momentum conservation
is simply realized by starting with renormalized classical functionals
$\CA_d \in F_d$, and by quantizing it with respect to the twist~$\CF_0$ and the map
$\tau_d$. On the other hand,
this quantiztion is equivalent to the conbined twist~$\CF_1$
(and the map $\tau$) from the point of view of the original functionals $\CA$.

Let us now focus on the symmetry.
Of course, there is the same kind of structure as (\ref{zero modes maps}) among
corresponding Hopf algebras.
It can be shown that the Poincar\'e--Hopf algebra $U(\CP)$ is a twist invariant
Hopf subalgebra of both $\CH_d$ and $(\CH_d)_{\CF_0}$.
This is proved in the same way as was done in \cite{AMW} for the twist $\CF_0$,
because only the following properties of the Green function $G$ are used:
$G^{\mu\nu}$ is proportional to $\eta^{\mu\nu}$, and
it is symmetric $G^{\mu\nu} (z,w)=G^{\nu\mu} (w,z)$.
Therefore, $U(\CP)$ remains algebraically
 a true symmetry under any step in (\ref{zero modes maps}).
On the other hand,
the violation of the momentum conservation in the original quantization
(\ref{original convention}) is due to the bad choice of a map $\tau$,
or equivalently the choice of a translation non-invariant vacuum state.

The argument in this subsection
seems to indicate that the dif\/feomorphism or the $B$-f\/ield background acquire
some ef\/fect from the extra normal ordering $\CN_d$ in the zero mode structure
and in the ef\/fective theory.
We hope to come back to this point in the future reference.

\section{Discussion}\label{section6}

As we have argued,
one of advantages of the twist quantization is that there is
a direct relation between quantization and symmetry
via the Hopf algebra.
One typical example is the Poincar\'e symmetry $U(\CP)$ which is
twisted by $\CF_1$ in a $B$-f\/ield background.
In order to compare the twist quantization with known formulations,
we formulated a useful method to decompose the twist into two successive twists.
It is then applied to several examples,
$B$-f\/ield background, f\/inite twisted dif\/feomorphism, and zero modes.

All the example of twist elements so far are coboundary, that are cohomologically trivial.
However, the decomposition is rather arbitrary and possible as far as
the second twist satisf\/ies the cocycle condition
under the f\/irst twist, the condition
corresponding to (\ref{2nd cocycle condition}).
Therefore, the method proposed in this paper can in principle be applied
to more non-trivial backgrounds,
such as non-constant metric or $B$-f\/ield.
For this, it is natural to ask whether the quantization in all the backgrounds
can be written as twist quantization, and if so, wheather the corresponding twist
elements are coboundary.
We think naively that any background corresponds
to a quantization with a coboundary twist,
since our star-products are the time ordered product, which is star-equivalent to
the ordinary commutative product,
but this has still to be proven.
The further study along this line
would help to get a better understanding of the structure
of the stringy geometry.
Another important and interesting open question is the relation between
a twist element for a generic background and the conformal symmetry.

\appendix

\section{Proof for successive twists}

We show two relations $(\Delta_{\CF_0})_{\CF_d}=\Delta_{\CF_1}$ and
$(S_{\CF_0})_{\CF_d} =S_{\CF_1}$ for successive twists (\ref{2step twist}).
The f\/irst equation is easily shown by def\/inition
$\Delta_{\CF_1}(h)=\CF_1 \Delta (h) \CF_1^{-1}
=\CF_d \CF_0 \Delta (h) \CF_0^{-1} \CF_d^{-1} =(\Delta_{\CF_0})_{\CF_d} (h)$
for $ \forall\,  h$.
To show the second equation, we recall that the def\/inition
$S_{\CF}(h)=US(h)U^{-1}$.
By using the notation $\CF=f^\alpha \otimes f_\alpha $ and
$\CF^{-1}=\bar{f}^\alpha \otimes \bar{f}_\alpha $,
operators $U, U^{-1} \in \CH$ is written as
$U=\mu ({\rm id}\otimes S)\CF=f^{\alpha}S(f_\alpha)$,
and similarly,
$U^{-1}=\mu (S\otimes {\rm id})\CF_{12}^{-1}=S(\bar{f}_{\alpha})\bar{f}^\alpha$.
We denote three $U$'s corresponding to $\CF_0$, $\CF_d$ and $\CF_1$ as
$U_0$, $U_d$, and $U_1$, respectively.
Then by using $\CF_1=\CF_d \CF_0$, we can rewrite $U_1$ as
$
U_1=\mu ({\rm id}\otimes S)\CF_1 =f_1^{\alpha}S(f_{1\alpha})
=f_d^{\alpha}f_0^{\beta }S(f_{0\beta }) S(f_{d\alpha})
=f_d^{\alpha}U_0 S(f_{d\alpha})
=\mu (1 \otimes U_0) ({\rm id}\otimes S)\CF_d
=[\mu (1 \otimes U_0) ({\rm id}\otimes S)\CF_d (1 \otimes U_0^{-1})] \cdot U_0
=[\mu ({\rm id}\otimes S_{\CF_0})\CF_d ] \cdot U_0
=U_d \cdot U_0
$.
Note that $U_d \in \CH_{\CF_0}$.
Therefore,
$
S_{\CF_1}(h)=U_1 S(h) U_1^{-1}
=U_d U_0 S(h) U_0^{-1} U_d^{-1}
=U_d S_{\CF_0} (h) U_d^{-1}
=(S_{\CF_0})_{\CF_d} (h)
$.

\subsection*{Acknowledgements}

The authors would like to thank M.~Mori for collaboration
and useful discussions. We also thank to Dr. U.~Carow-Watamura for useful comments and
discussions.
This work is supported by Grant-in-Aid for Scientif\/ic Research from the Ministry of Education, Culture, Sports, Science and Technology, Japan, No. 19540257.

\pdfbookmark[1]{References}{ref}
\LastPageEnding

\end{document}